# Exploring AI Capabilities in Participatory Budgeting within Smart Cities: The Case of Sao Paulo

Presented at 28th IPSA World Congress of Political Science, Seoul 2025.

Panel: RC05.03 - Smart Cities, AI, Online Participation, and Local State Capacity

Italo Alberto Sousa[1]

Mariana Carvalho da Silva[2]

Jorge Machado[3]

José Carlos Vaz[4]

**Abstract**

This research examines how Artificial Intelligence (AI) can improve participatory budgeting processes within smart cities. In response to challenges like declining civic participation and resource allocation conflicts, the study explores how online political participation can be improved by AI. It investigates the state capacity governments need to implement AI-enhanced participatory tools, considering technological dependencies and vulnerabilities. It analyzes technological and administrative structures, actors, interests, and strategies to understand the dynamics of online political participation technologies in the case of Sao Paulo, Brazil. The study contributes to understanding how technological advancements can reshape participatory budgeting processes. In a broader sense, the research highlights how AI can transform participatory institutions by offering new tools for citizens and also for government officials in charge of participatory processes within smart cities.

**Keywords**

[1]University of Sao Paulo, School of Arts, Sciences and Humanities, Brazil
[2]University of Sao Paulo, School of Arts, Sciences and Humanities, Brazil
[3]University of Sao Paulo, School of Arts, Sciences and Humanities, Brazil
[4]University of Sao Paulo, School of Arts, Sciences and Humanities, Brazil; Brazil Chair at the University of Münster, Institute for Plitical Science, Germany

Artificial Inteligence, Online Participation, State Capacity, Smart Cities, Digital Participatory Budgeting

**Introduction**

The debate on the use of Artificial Intelligence (AI) in online social participation in smart cities is at the heart of discussions on the future of digital democracy and urban governance. The promise that AI can expand the reach, efficiency, and quality of participatory processes coexists with concerns about risks, inequalities, and ethical challenges (Campagnucci et al., 2025; OECD, 2020). Beyond techno-solutionist enthusiasm and alarmist critiques, it is essential to critically and substantively analyze the real possibilities of these technologies and, above all, the new demands they impose on state capacities—particularly in the context of smart cities.

To begin reflecting on this phenomenon, this research adopted a case study strategy to identify potential applications of AI tools in online participatory processes. The starting point was the identification of existing problems and limitations, along with the increasingly available AI resources, thereby highlighting the demands for new state capacities.

In this sense, the case of the digital participatory budgeting initiative in the city of São Paulo presents an excellent opportunity to analyze both the potential for AI adoption and the state capacities required for its effective implementation. This initiative uses an internationally recognized platform (CONSUL) and is implemented in a major Global South metropolis (the city of São Paulo has over 12 million inhabitants), in a country with a tradition of social participation and participatory budgeting that was an international reference in the 1990s and 2000s but has experienced a significant decline in adoption and engagement in participatory budgeting processes in recent decades (Bezerra & Junqueira, 2022).

Following this introduction, the article systematizes, based on selected literature, potential uses of AI in online participatory processes in smart cities. It then further explores the literature on state capacities and AI. The adopted methodology is described in the next section. The subsequent section presents a characterization and analysis of the studied case, providing the basis for the following section, which discusses prospects for AI use in São Paulo's Citizen Budget.

**Possibilities of AI Use in Online Social Participation**

In the context of smart cities, AI should be understood as part of a sociotechnical infrastructure—that is, a set of systems, data, processes, actors, and power relations that intertwine and shape the possibilities and limits of online social participation (Crawford, 2021; Campagnucci et al., 2025). It is not merely about algorithms but an ecosystem in which political decisions, economic interests, and social values influence the design, implementation, and use of technologies. In this ecosystem, technology is the product of a social construction involving actors' strategies, their interests and networks, as well as the cognitive frameworks of decision-makers and users (Vaz, 2017).

The literature highlights that the adoption of AI for social participation occurs amid disputes over proprietary versus open-source models, the centralization of technological power in large corporations, and tensions between innovation and regulation (Campagnucci et al., 2025; Morozov, 2013). In this scenario, the definition of the data used, the transparency of algorithms, and the governance of systems become central political issues, directly impacting the legitimacy and effectiveness of participatory processes (European Commission, 2022).

The application of AI in online social participation platforms can be grouped into four key political functions: analysis and systematization of demands, facilitating dialogue and deliberation, inclusion and accessibility, visualizing realities and futures (Sousa, Machado & Vaz, 2025; OECD, 2020).

Analysis and systematization of demands : AI can be employed in analyzing large volumes of data generated by public consultations, participatory budgeting, and deliberation platforms. Natural language processing algorithms and topic modeling can identify patterns, cluster demands, and extract summaries of opinions, making public listening processes more efficient (Jungherr, Posegga & An, 2020; OECD, 2020). Platforms like Pol.is, used not only in Brazil, exemplify the potential of these tools for mapping consensus and disagreements (Pol.is, 2024).

Facilitation of dialogue and deliberation: Another dimension where AI can contribute is in facilitating dialogue. Clustering algorithms and automated moderation can support large-scale debates by suggesting topics, identifying points of convergence, and promoting constructive interactions (Smith & Setälä, 2018). However, studies indicate that algorithmic mediation can also reinforce polarization and limit plurality if not carefully designed (Jungherr, Posegga & An, 2020).

Inclusion and accessibility: AI can enhance inclusion and accessibility through recommendation systems, accessible chatbots, automatic translation, and language simplification, reducing barriers for traditionally underrepresented groups (World Economic Forum, 2021; OECD, 2020). These solutions are particularly relevant in contexts of digital inequality, where access to and understanding of platforms remain significant challenges (Eubanks, 2018).

Visualization and simulation tools: AI-based visualization and simulation tools, such as urban digital twins and augmented reality systems, allow citizens to visualize future scenarios and simulate the impact of different public policies (Batty et al., 2012). By translating complex data into accessible representations, these technologies facilitate the understanding of urban challenges and encourage informed participation (World Economic Forum, 2021).

Despite these possibilities, the literature is replete with warnings about the risks and ethical dilemmas associated with AI use in social participation. Key challenges include algorithmic biases, opinion manipulation, amplification of extremist discourse, exclusion of minorities, and opacity in decision-making processes (O'Neil, 2016; Mittelstadt et al., 2016; Campagnucci et al., 2025).

The opacity of algorithms—often protected by trade secrets—hinders public auditing and undermines accountability (European Commission, 2022). Moreover, the centralization of technological power in a few companies limits governments' digital sovereignty and may create dangerous dependencies (Morozov, 2013; Crawford, 2021).

Another critical issue is the tendency toward automating normative decisions, which cannot be reduced to mathematical solutions, as they involve values, conflicts, and collective deliberation. The risk of "technological solutionism" is real and may lead to the depoliticization of participatory processes (Morozov, 2013).

**AI Use in Online Participation and New Demands on State Capabilities**

The concept of state capacity refers to the set of competencies, resources, and institutional structures that enable the state to formulate, implement, and evaluate public policies effectively (Fukuyama & Levy, 2010; Mazzucato, 2018). In the context of smart cities, these capacities take on specific contours, as they involve the ability to handle advanced digital technologies, manage large volumes of data, and coordinate multiple actors in dynamic and complex environments (Janssen, Charalabidis & Zuiderwijk, 2012).

The particularity of technological capacities lies in the state's need not only to acquire ready-made solutions but also to develop autonomy and digital sovereignty (Kattel & Mazzucato, 2018). In the case of Artificial Intelligence, this means being able to critically evaluate technological options, opt for open-source solutions where possible, ensure control over data and algorithms, and avoid excessive dependence on private vendors (Campagnucci et al., 2025; OECD, 2019).

The adoption of AI for social participation requires the state to possess technical competencies for the selection, implementation, customization, and monitoring of systems. This includes everything from analyzing security and privacy requirements to the ability to audit algorithms and ensure interoperability between different platforms (European Commission, 2022).

The literature emphasizes the importance of multidisciplinary teams that combine expertise in data science, political science, law, ethics, and communication (Sousa, Machado & Vaz, 2025; OECD, 2020). Continuous training for public servants is essential to keep pace with rapid technological evolution and mitigate risks of obsolescence or misuse of tools (United Nations, 2021; Sousa, Machado & Vaz, 2025).

Beyond technical competencies, institutional capacities are needed to design participatory processes that incorporate AI in an inclusive and transparent manner (Smith, 2009; Fung, 2015). This entails creating clear regulations, defining transparency and accountability standards, establishing mechanisms for social oversight, and ensuring the effective participation of different societal segments in the governance of these technologies (Open Government Partnership, 2023).

Algorithmic governance, as highlighted by the European Commission (2022), should be guided by ethical principles such as fairness, explainability, non-discrimination, and respect for human rights. The adoption of open-source models, the publication of data in accessible formats, and detailed documentation of processes are recommended strategies to strengthen public trust (Pol.is, 2024; OECD, 2019).

Accountability is a central aspect of AI adoption for social participation. Public managers must bear ultimate responsibility for decisions and outcomes generated by automated systems, without shifting blame to technology (Bovens, 2007; Mittelstadt et al., 2016). This requires constant oversight, community engagement, and transparency regarding the implementation and impacts of adopted tools.

The participation of civil society and experts in the evaluation and monitoring of systems is crucial to ensuring democratic principles are upheld (Campagnucci et al., 2025; OECD, 2020). The literature recommends establishing advisory councils, independent audits, and mechanisms for reporting and correcting errors (European Commission, 2022; Open Government Partnership, 2023).

Several authors highlight the need for specific state capacities to mitigate risks to democracy posed by AI adoption in online participatory processes. These include: (a) Continuous evaluation and monitoring, with systematic bias testing, result validation, and technical and social audits of systems (Mittelstadt et al., 2016; European Commission, 2022); (b) open transparency practices, data and algorithmic governance, prioritizing open-source solutions and publishing data in accessible formats (OECD, 2019; Pol.is, 2024, Machado, 2025); and (c) resources for including vulnerable and underrepresented groups, particularly those affected by digital exclusion due to low digital literacy. This requires maintaining hybrid (online and offline) participation channels, investing in digital training, and ensuring accessibility for marginalized groups (World Economic Forum, 2021; Eubanks, 2018).

These capacities are directly linked to capabilities for fostering civil society engagement—not only in participatory processes but also in the governance of AI-based technological solutions. This requires encouraging the participation of social organizations, universities, and experts in the governance of technologies, from design to system evaluation (Campagnucci et al., 2025; Fung, 2015).

**Methodological Approach**

Having determined the strategy of analyzing the problem through a case study, the case of São Paulo's *Orçamento Cidadão* (Citizen Budget) was selected. To characterize and analyze the case, a qualitative and descriptive approach was adopted, based on the application of Vaz's (2024) analytical framework on state capacities for online participation.

Official data were obtained through document analysis, examining official records from the São Paulo City Hall, including ordinances (e.g., SF No. 126/2024), decrees (e.g., No. 64,017/2025), management reports available on the *Participe Mais* platform, and publications from the 2nd Open Government Action Plan. These materials allowed for the reconstruction of the legal framework, process stages, and quantitative data on participation and budget execution (São Paulo, 2025).

The platform used in the initiative underwent structured inspection to understand its available functionalities. Semi-structured interviews with key actors—public servants involved in managing the Orçamento Cidadão—were conducted, covering topics such as operational challenges, technology use, and interaction with civil society. These interviews enabled data triangulation and supplemented information missing from official documentation. They also provided critical insights into the initiative's gaps and difficulties.

Based on these analyses, the problems, limitations, and potential expansions of the initiative were contrasted with the possibilities outlined in the literature on AI applied to online participation. This comparison identified feasible applications of AI and the state capacities required by the municipal government for their effective implementation.

**Characterization of the Digital Participatory Budget in São Paulo**

The Citizen Budget of the city of São Paulo represents an experiment in digital democracy in Brazil, integrating civil society demands into the municipal budget cycle through an online platform and in-person public hearings. Implemented as part of the 2nd Open Government Action Plan (2018-2020), the process replaced the traditional model of public hearings with a hybrid format, combining digital interactions and face-to-face meetings. The Participe Mais platform, developed in partnership with UNESCO and based on the free software CONSUL (originally created by the Madrid City Council), serves as the main participation channel. Below, the initiative is described and analyzed, using as reference the model established by Vaz (2024). The categories proposed in the model were adapted for this study.

*Institutional Context and Legal Framework*

São Paulo is the largest Brazilian city, with about 21 million inhabitants in its metropolitan area. Approximately 12 million are residents of the municipality of São Paulo, which covers a varied territory spanning 1,521 km². The municipality is divided into 32 administrative districts (subprefeituras), which have low administrative autonomy, focusing more on urban maintenance activities and articulating local demands to be addressed by the centralized structures of the municipal government.

Most of the municipality's territory is densely urbanized, with significant socioeconomic disparities observed between its districts and within them. Areas of intense real estate appreciation, high-income residents, abundant infrastructure, and a high concentration of qualified jobs contrast with others characterized by precarious or low-quality housing conditions, deficiencies in infrastructure and public services, and few economic opportunities for their residents. Over time, residents of the poorest areas have been underrepresented in political participation processes, in contrast to middle-class sectors and the economic elite, who have greater access to state decisions and the press. This scenario generates a high level of demands on the municipal government, which are not always adequately addressed.

The Citizen Budget of the municipality of São Paulo is an initiative that allows citizens to submit proposals for inclusion in the annual municipal budget and to express their opinions on them through non-binding voting. It was formalized through a series of normative instruments that establish its guidelines and guarantee specific resources for its execution. Ordinance SF No. 126/2024 establishes the participatory process, while Decree No. 64.017/2025 allocates R$10 million per administrative district for the implementation of approved proposals, totaling R$320 million annually (São Paulo, 2025). This measure sought to ensure that popular demands were effectively incorporated into the budget, although criticisms have emerged regarding the equal distribution of resources, which does not consider the socioeconomic disparities between the city's regions.

Additionally, the process aligns with the Budget Guidelines Law (LDO), which establishes popular participation as one of the fundamental principles of budget formulation. The LDO reinforces the need for transparency and social control, ensuring that citizens can monitor all stages of the budget cycle (São Paulo, 2025).

*Scope and Political Functionalities*

The main political functionalities presented by the Citizen Budget are proposal submission, online voting, and tracking the status of demands. The presentation and voting of proposals are non-binding, meaning the city government may or may not accept citizens' suggestions.

The Citizen Budget of the São Paulo City Hall follows a structured process consisting of eight interconnected phases, combining in-person and digital participation. These stages were designed to ensure transparency, technical feasibility, and citizen engagement in the allocation of the R$320 million annually allocated to the program. Based on official documents and accounts from managers, each phase of its latest edition can be detailed as follows:

*Phase 1: Proposal Submission (April-May)*

The initial phase occurs simultaneously in two formats. In the in-person format, public hearings are held across all 32 administrative districts, where citizens present demands through thematic working groups. In 2025, 2,414 proposals were collected, with 60% originating from these events, according to one interviewee.

In the digital format, the platform allows registration of individual or collective proposals, requiring personal data and thematic categorization (e.g., mobility, health). Data shows 40% of proposals came through this channel, with higher participation from central districts (São Paulo, 2025).

*Phase 2: Prioritization by Participatory Bodies (May-June)*

Municipal Participatory Councils (CPMs) - elected by residents in each district (subprefeitura) - screen demands and select 10 popular proposals per district based on criteria including alignment with Regional Plans and number of platform endorsements.

Additionally, CPMs develop 5 technical proposals incorporating longstanding unaddressed demands. When CPMs fail to fulfill this role, DIPAR's team assumes prioritization per Decree No. 64.017/2025.

*Phase 3: Technical Feasibility Assessment (June-July)*

The 15 prioritized proposals per region undergo evaluation by technical bodies. Municipal secretariats analyze legal and budgetary compliance (average 20-day review period), while districts assess local implementability. In 2025, 30% of proposals were deemed unfeasible, primarily due to conflicts with existing sectoral plans (interviewee data).

*Phase 4 - Appeal (July)*

The appeals phase represents a crucial democratic innovation in the Citizen Budget, implemented in 2025 to expand social oversight over technical decisions. In this stage, the Municipal Participatory Councils (CPMs) in each *subprefeitura* (districts) can challenge feasibility analyses they deem inadequate by completing a structured form on the Participe Mais platform within five business days. Appeals must contain detailed technical or legal justifications, preferably accompanied by supporting documentation. In the 2025 edition, an average of 6.8 appeals per district were observed. Of the 217 appeals filed, 18% resulted in revised decisions. However, the phase revealed significant challenges: the concentration of 73% of appeals in regions with more structured CPMs highlights inequalities in civil society's capacity for articulation, while DIPAR's reduced staff (only seven civil servants) proved insufficient to thoroughly analyze all appeals within the stipulated timeframe.

*Phase 5 - Public Voting (August-September)*

The public voting, reorganized in 2025 to occur after the feasibility analysis, constitutes the moment of greatest direct public engagement. Using a preferential voting system on the Participe Mais platform, each citizen can rank up to five proposals per district, with automatic allocation of resources until the R$10 million available per region is exhausted. The schedule change reduced by 40% the frustration with elected proposals that were later deemed unfeasible. Geographic analysis reveals distinct patterns: peripheral regions, with an average of 320 votes per district, prioritized basic sanitation (38%) and public lighting (29%), while the expanded center, with only 85 votes on average, focused on active mobility (51%) and cultural facilities (23%). However, the predominance of desktop access (92%) over mobile devices (8%) and the concentration of 61% of votes in the final three days point to technological and cultural barriers that limit participation.

*Phase 6 - Incorporation into the Annual Budget Bill (September)*

The transition of popular proposals to the Annual Budget Bill (PLOA) requires meticulous technical work by the Division of Participatory Processes. Each citizen demand is reformulated in standard budgetary language - for example, "build daycare center" becomes "expansion of daycare network" - with specification of physical targets and corresponding financial allocation. In 2025, 83% of proposals were allocated to the "Urbanization" (10.122) and "Basic Health Care" (08.141) functions. This technical translation process generates frequent tensions, particularly in cross-cutting proposals involving multiple secretariats - approximately 12% of cases require additional legal drafts to resolve jurisdictional conflicts. The phase also reveals the delicate balance between popular will and administrative feasibility, when highly voted projects need to be adapted to constraints of existing sectoral planning.

*Phase 7 - Feedback Hearings (October)*

The feedback hearings, conducted in hybrid format (in-person/virtual), face growing challenges of legitimacy and effectiveness. Divided into two segments - technical presentation of analyses (45 minutes) and open microphone (2 minutes per participant) - these sessions attracted only 12% of participants from previous phases in 2025. The participant profile reveals serious imbalances: 78% were male and 65% were over 50 years old. Furthermore, 41% of interventions deviated from budgetary scope, addressing demands unrelated to the process. The absence of specialized technicians in 60% of districts exacerbated the problem, resulting in superficial explanations of decisions. Innovation attempts, such as interactive chats and explanatory infographics, have so far failed to reverse the growing disconnect between government and citizens in this crucial accountability phase.

*Phase 8 - Monitoring (Following Year)*

The monitoring system represents the most fragile link in the participatory cycle. Based on quarterly updates by implementing agencies in the Strategic Monitoring and Follow-up System (SMAE), supplemented by sample checks by the Comptroller's Office, the mechanism shows alarming discrepancies. Data from 2024 reveals that while reported financial execution reached 89% for maintenance works, on-site inspections confirmed only 37% effective implementation. Approximately 22% of proposals remained without updates for over eight months, undermining process transparency.

The lack of integration with the Budget Management System (SIGOR) and absence of automatic notifications for registered citizens severely limit the potential for social oversight in this phase. These gaps partially explain the growing public distrust in the concrete effectiveness of the Citizen Budget, despite progress in other process stages.

Engagement in the Citizen Budget has shown significant variations over the years, with a declining trend in the number of proposals and votes. According to data available on Participe Mais, the 2021 edition recorded 3,519 proposals and 43,000 votes, while in 2025 these numbers dropped to 2,414 proposals and 6,070 votes (São Paulo, 2025). This reduction represents a decline of approximately 31% in the volume of proposals and 86% in the number of votes over five years.

This low participation becomes even more evident when compared to São Paulo's population, which exceeds 11.45 million inhabitants. In 2025, for example, only 0.05% of citizens participated in the process, raising questions about representativeness and the effectiveness of outreach strategies. Among the factors that may explain this phenomenon are distrust in the process's effectiveness, lack of access to digital tools by marginalized groups, and insufficient publicity by the government, according to one interviewee.

Starting in 2025, the Citizen Budget was allocated a fixed budget of R$10 million per district, totaling R$320 million annually. These funds are exclusively dedicated to implementing approved proposals, representing progress compared to previous editions when there was no guarantee of specific resources (São Paulo, 2025).

However, this measure has also generated tensions. The equal distribution of resources does not account for territorial inequalities, potentially perpetuating disparities between more and less developed regions. Additionally, proposals exceeding R$10 million are automatically deemed unfeasible, limiting the capacity to address structural issues. Another controversial point is the use of remaining funds, which are reallocated by Municipal Participatory Councils (CPMs) without a transparent accountability process, as reported by one interviewee.

The proposals submitted by the population primarily focus on urban maintenance themes, such as public lighting, cleaning, and sidewalk maintenance, as well as local infrastructure, including the construction or renovation of parks, schools, and health centers. Demands related to urban mobility, such as bike lane implementation and public transportation improvements, also frequently appear (São Paulo, 2025).

Despite this diversity, the process has been criticized for not addressing broader structural issues, such as social inequality, housing, and public safety. This limitation partly reflects the institutional design of the Citizen Budget, which prioritizes smaller-scale, quickly implementable interventions over transformative policies.

*Actors Involved and Their Roles*

The Citizen Budget involves a complex network of actors, each with specific responsibilities. The Division of Participatory Processes (DIPAR), linked to the Executive Secretariat of Planning (SEPLAN), is primarily responsible for the overall coordination of the process, including organizing public hearings, monitoring proposals, and liaising with other municipal agencies.

Municipal Secretariats and District Administrations play a central role in analyzing the technical, legal, and budgetary feasibility of proposals. However, interviews with civil servants reveal that many agencies struggle to meet established deadlines, and analyses are not always sufficiently detailed to enable productive dialogue with the public.

The Municipal Participatory Councils (CPMs), elected by civil society, are tasked with prioritizing demands from each district. Despite their importance, many council members lack adequate training, which may compromise the quality of proposal selection. Furthermore, the absence of an institutionalized process for the participation of marginalized groups—such as favela residents and the homeless population—limits the process's representativeness (ENGELBERT et al., 2021).

Finally, civil society participates by submitting proposals and voting online, but low engagement suggests the process has yet to broadly engage citizens. Qualitative data indicate that participation is mostly individual, with little involvement from organized groups—a contrast to São Paulo's historical tradition of social mobilization (KUBICEK and AICHHOLZER, 2016).

*Technology Used*

The "Participa Mais" platform, based on the CONSUL software, serves as the primary digital tool for the Citizen Budget. Developed in partnership with UNESCO, the system enables user registration (requiring personal data), proposal submission, online voting, and tracking of demand status.

Despite its basic functionality, the platform presents significant limitations. For instance, it lacks an analytical dashboard to visualize consolidated data on participation or budget execution. Furthermore, the absence of integration with social networks or mobile applications restricts its reach, particularly among youth and peripheral populations who predominantly rely on smartphones for internet access (CHATFIELD et al., 2015).

*Data Protection and Usage*

The Citizen Budget follows guidelines from Brazil's General Data Protection Law (LGPD), collecting a limited set of personal information including full name, address, and personal ID numbers from participants. However, there is no transparency regarding how this data is used to inform public policies or improve the platform. Additionally, proposal-related information is not made available in open formats, hindering its reuse by researchers and civil society organizations.

*Impact and Effectiveness*

Participation in the Citizen Budget has shown significant variations over the years, with a declining trend in both the number of proposals and votes. According to data available on Participe Mais, the 2021 edition recorded 3,519 proposals and 43,000 votes, while in 2025 these numbers decreased to

2,414 proposals and 6,070 votes (São Paulo, 2025). This reduction represents a decline of approximately 31% in the number of proposals and 86% in votes over five years. Table 1 below shows the evolution of citizen participation in the process.

**Table 1: Evolution of Citizen Participation in the São Paulo Citizen Budget**

|  | 2021 | 2022 | 2023 | 2024 | 2025 |
|---|---|---|---|---|---|
| **Number of Collected Proposals** | 3.519 | 3.130 | 2.023 | 3.313 | 2.414 |
| **Commitments Made by Agencies** | 70 | 84 | 89 | 111 | 83 |
| **Elected Proposals** | 163 | 162 | 170 | 192 | 116 |
| **Number of Votes** | 43.000 | 24.578 | 12.005 | 15.328 | 6.070 |

Source: Authors' own elaboration with data extracted from the platform.

The low participation becomes even more evident when compared to São Paulo's population, which exceeds 11.45 million inhabitants. In 2025, for example, only 0.05% of citizens participated in the process, raising questions about representativeness and the effectiveness of outreach strategies. Among factors that may explain this phenomenon are distrust in the process's effectiveness, lack of access to digital tools by marginalized groups, and insufficient publicity by the government, according to one interviewee.

The Citizen Budget has brought important advances in terms of transparency, since all proposals and feasibility analyses are public and accessible on the platform. The guarantee of specific resources for approved demands (R$320 million/year) also represents progress compared to previous models, where proposals competed with other budgetary priorities (São Paulo, 2025).

Another positive aspect is the integration between municipal agencies, which began working in a more coordinated manner on intersectoral proposals. For example, a demand for tree planting may involve both the Environment Secretariat and the local District Administration, encouraging more collaborative management.

Low popular participation remains the biggest obstacle to the process's success. As seen, the number of votes dropped drastically between 2021 and 2025, indicating a potential legitimacy crisis. Moreover, the focus on localized interventions, such as sidewalk and square repairs, limits the transformative potential of the Citizen Budget, which contributes little to reducing structural inequalities (Weiss, 2023).

The lack of systematic evaluation is another problem. There are no consolidated reports comparing proposal implementation across years, making it difficult to identify patterns and correct flaws.

Monitoring, when it exists, is fragmented and depends on individual access to each proposal on the platform, without specific functionalities for this purpose.

**Perspectives for the Use of AI in São Paulo's Citizen Budget**

From the analysis of the functioning and trajectory of São Paulo's Citizen Budget and the examination of literature and international experiences on the use of AI for citizen participation, we have identified five main axes of challenges and opportunities for improving the initiative. The study reveals that, although the program has the potential to contribute to the democratization of public budgeting, it faces structural obstacles that limit its transformative potential, many of which could be mitigated through the strategic application of Artificial Intelligence technologies.

The identified problems concentrate on five critical dimensions: (1) low participation and inequalities in citizen engagement, combined with evident digital exclusion of peripheral populations; (2) the high rate of technical unfeasibility of proposals, resulting from disconnects between popular demands and administrative possibilities; (3) weaknesses in monitoring and accountability, which undermine the credibility of the process; (4) difficulties citizens face in formulating technically consistent proposals; and (5) technological limitations of the digital platform, which does not incorporate more recent developments, especially in Artificial Intelligence.

The possibilities for expansion using AI emerge as promising paths to overcome these challenges. Solutions such as predictive engagement analysis, virtual assistants for proposal drafting, automated pre-feasibility analysis systems, and intelligent monitoring tools offer potential to significantly improve all stages of the process. Particularly relevant is AI's ability to serve as a bridge between the technical knowledge of public administration and citizens' everyday demands, reducing information and power asymmetries.

However, the successful implementation of these innovations is conditioned on the development of specific state capacities. International experience demonstrates that AI projects in public administration require: robust infrastructure of open and interoperable data; multidisciplinary teams capable of mediating between technology and citizenship; transparent mechanisms of algorithmic governance; and continuous investments in digital inclusion. The São Paulo case suggests that combining these elements could transform the Citizen Budget into a global reference for smart participatory democracy.

Following this, each of these critical dimensions is analyzed, presenting the problems, limitations, and possibilities for expansion, supplemented by the potential contributions of Artificial Intelligence and the state capacities required for its implementation.

*Dimension 1: Expanding Engagement and Reducing Participatory Inequalities*

One of the main challenges of São Paulo's Citizen Budget is the low popular participation, which has shown a sharp decline in recent years. The data reveal that while 43,000 votes were recorded in 2021, this number dropped to just 6,070 in 2025 - an 86% reduction. When compared to São Paulo's population of over 11 million inhabitants, this participation represents only 0.05% of citizens, raising serious questions about the process's representativeness.

In addition to the overall low engagement, participant profiles show marked inequalities. Central districts and wealthier areas of the city have significantly higher participation than peripheral regions. This disparity reflects structural barriers, such as digital exclusion - evidenced by the fact that 92% of accesses to the Participe Mais platform occur via computers, while only 8% come from mobile devices, precisely the main means of internet access in peripheral areas. Other factors contributing to this asymmetry include the lack of targeted outreach and distrust in the process's effectiveness among historically marginalized populations.

Artificial Intelligence could significantly contribute to addressing these challenges. Through predictive data analysis, it would be possible to identify participation patterns and cross-reference them with socioeconomic indicators, enabling more targeted mobilization campaigns for underrepresented areas. Chatbot systems integrated with popular applications like WhatsApp could send personalized notifications and explain the process in accessible language, increasing engagement among groups less familiar with digital tools. Furthermore, recommendation algorithms could suggest more relevant content and proposals for each user profile, increasing interest in participation.

To implement these solutions, the city government would need to develop specific institutional capacities. The first would be integrating different municipal databases, such as social registries, income indicators, and digital infrastructure maps, creating a comprehensive overview of participatory inequalities. Another crucial aspect would be establishing partnerships with telecom operators to reduce mobile access costs in peripheral regions, ensuring digital solutions truly reach the entire population. Finally, it would be necessary to invest in multidisciplinary teams capable of interpreting AI-generated data and translating it into effective engagement strategies, combining technical knowledge with an understanding of the city's territorial dynamics.

*Dimension 2: Improving Technical Feasibility and Quality of Proposals*

One of the main points of tension in São Paulo's Citizen Budget concerns the high rate of technical unfeasibility of proposals submitted by the population. Data from 2025 reveal that approximately 30% of submitted demands were considered unfeasible by municipal technical bodies, generating frustration among participants and raising questions about the process's effectiveness. This situation stems from multiple factors, including the disconnect between citizens' expectations and the budgetary and legal possibilities of public administration, as well as participants' difficulty in accessing information about existing sectoral planning.

The appeals phase, created specifically to review these technical decisions, has proven to be ineffective in practice, with only 18% of reviews being approved. This low reversal rate of technical analyses indicates that, in many cases, the proposals are indeed incompatible with the city's urban planning or budgetary guidelines, but it also reveals the need to improve the quality of proposals from their initial formulation. The situation is worsened by the fact that many citizens are unaware of basic parameters for developing viable demands, such as reference values for public works or the specific responsibilities of each municipal secretariat.

In this context, Artificial Intelligence offers promising solutions to enhance the process of drafting and analyzing proposals. Automated pre-analysis systems could cross-reference submitted demands with databases containing master plans, budgetary laws, and previously executed projects, providing immediate feedback on potential incompatibilities. Natural language processing algorithms could assist in translating colloquial demands into the technical language appropriate for administrative processes, reducing communication gaps. Additionally, machine learning-based recommendation tools could suggest adjustments to proposals, indicating viable alternatives when the original version conflicts with existing planning.

The implementation of these solutions would require the municipal administration to develop specific institutional capacities. First, it would be necessary to structure a comprehensive process of digitizing and organizing municipal technical archives, including urban laws, previous executive projects, and legal opinions, creating databases accessible to AI systems. Simultaneously, it would be essential to train public servants to act as mediators between automated technical analyses and citizens, ensuring that the systems' recommendations are understandable and useful. Another fundamental aspect would be establishing clear protocols for algorithmic governance, ensuring transparency in the criteria used by AI tools and mechanisms for appeal when citizens disagree with automated assessments.

### *Dimension 3: Strengthening Transparency and Monitoring of Approved Proposals*

The Citizen Budget cycle faces serious deficiencies in its final monitoring and accountability phase, undermining the credibility of the entire process. Data from 2024 reveals an alarming discrepancy between official reports and reality: while 89% of projects were marked as "completed" in municipal systems, on-site inspections confirmed only 37% actual implementation. This gap between recorded and actual progress stems from multiple weaknesses, including the lack of integration between management systems (SMAE and SIGOR), the absence of automatic citizen notifications, and insufficient systematic verification by oversight bodies.

The situation is exacerbated by the fact that 22% of proposals remain without updates for over eight months in the monitoring system, leaving participants uninformed about the progress of demands they helped elect. This opacity in the implementation stage fosters growing distrust in the Citizen Budget's actual effectiveness, as evidenced in feedback hearings where only 12% of previous phase participants attended, demonstrating the weakening of this crucial accountability stage. Furthermore, the predominance of superficial interventions in these hearings (41% of statements deviated from budgetary scope) indicates poor-quality dialogue about policy implementation.

Artificial Intelligence offers powerful tools to overcome these challenges and build a robust monitoring system. Natural Language Processing (NLP) techniques could automatically extract deadlines, targets, and implementation stages from technical reports, generating alerts for delays and inconsistencies. Computer vision systems could analyze georeferenced photos submitted by inspectors or citizens, comparing "before and after" images to verify actual project progress. Additionally, predictive analysis algorithms could identify chronic delay patterns by region or project type, enabling preventive interventions by the administration.

Effective implementation of these solutions would require strengthening key state capacities. First, it would be necessary to develop a robust open data infrastructure with APIs enabling real-time integration between various municipal systems. Second, establishing strategic partnerships with oversight bodies, such as the Municipal Comptroller General, would be essential for cross-validating information generated by automated systems. Finally, creating algorithmic governance mechanisms would be crucial to ensure auditability of AI systems, guaranteeing their analyses and recommendations remain transparent and contestable.

*Dimension 4: Democratizing Proposal Development with Technological Support*

The proposal formulation process in São Paulo's Citizen Budget presents structural challenges that limit citizens' ability to submit technically sound demands. Data analysis reveals that a significant portion of proposals are drafted in generic or incomplete ways, hindering their evaluation and implementation. This situation stems from multiple barriers faced by participants, including unfamiliarity with budgetary technical language, difficulty accessing urban planning information, and lack of familiarity with municipal administrative processes. The problem worsens when considering the educational and digital inequalities present in São Paulo's population, which create asymmetries in effective participation capacity.

The complexity of the proposal development process has concrete consequences for participation quality. Many demands are presented vaguely (such as "improve healthcare" or "more security"), lacking minimum required elements for technical analysis like precise locations, proper scaling, or cost estimates. Other well-intentioned proposals encounter legal or budgetary issues that could have been identified beforehand. This creates a dual problem: citizens become frustrated when their ideas are rejected, while the municipal administration spends considerable effort analyzing proposals that are fundamentally flawed from inception.

The strategic application of Artificial Intelligence could transform this scenario by democratizing access to proposal development tools. Generative AI systems could act as intelligent assistants, helping citizens transform initial ideas into well-structured proposals. These tools could suggest missing elements, alert about legal requirements, and even provide rough cost estimates based on similar executed projects. Another promising application would be developing intelligent databases that cross-reference draft proposals with urban planning information, flagging potential conflicts with master plans or ongoing projects. Additionally, automated translation systems could make the technical language of notices and regulations more accessible, reducing information asymmetry between citizens and administration.

Successful implementation of these solutions would require developing specific state capacities. First, it would be necessary to structure an ecosystem of open and interoperable data, integrating information about budgets, urban planning, and past bids. Second, investing in training public servants to mediate between technological systems and citizens would be essential to ensure proper and inclusive tool usage. Finally, establishing technological governance mechanisms would be crucial to guarantee algorithm transparency and the possibility of human review of automated suggestions.

### *Dimension 5: Modernizing the Digital Platform and Participatory Inclusion*

The Participe Mais platform, the main digital interaction channel for the Citizen Budget, has technological limitations that undermine its potential as a democratic participation tool. Data shows only 8% of accesses come from mobile devices, despite smartphones being the primary internet connection method in peripheral areas. This discrepancy reveals a serious access barrier that excludes significant population sectors from the participatory process. Additionally, the platform lacks basic accessibility features like foreign language translation tools or adaptations for visually impaired users, further limiting its democratic reach.

The platform's technological obsolescence manifests in various functional aspects. The interface has complex information architecture with unintuitive navigation and overly technical terminology. The absence of analytical dashboards prevents citizens from clearly viewing proposal progress or participation history in their region. Lack of integration with social media and messaging apps restricts content virality and organic mobilization. These limitations create a frustrating user experience, particularly for those less digitally familiar, contributing to the observed low engagement rates.

Artificial Intelligence offers transformative solutions for these challenges. Adaptive design systems could automatically reconfigure the interface based on access devices, creating smartphone-optimized versions without dedicated apps. Content personalization algorithms could adapt language and information detail based on user profiles, making the platform accessible to both laypeople and experts. AI-based accessibility tools, like automatic audio transcription to sign language or image descriptions for visually impaired users, could significantly expand the reached audience. Moreover, intelligent recommendation systems could connect citizens with similar interests, fostering collaborative networks around common proposals.

Implementing these innovations would require developing specific state capacities. It would be necessary to establish a permanent multidisciplinary team for platform management, combining technology experts, user experience designers, and public managers. Development should adopt open-source principles, allowing public auditing and reuse by other cities. A continuous usability testing program with diverse groups would ensure solutions meet actual population needs. Finally, creating structured feedback channels would enable iterative platform improvements based on real user experiences.

**Final Considerations**

The analysis presented in this article demonstrates that the strategic application of Artificial Intelligence holds transformative potential for São Paulo's Citizen Budget, offering innovative solutions to its main challenges. AI can serve as a powerful tool for democratic equalization, reducing historical participation asymmetries through: (1) predictive systems to expand engagement in peripheral areas; (2) virtual assistants that democratize the development of viable proposals; (3) automated monitoring mechanisms that strengthen transparency; and (4) adaptive platforms that overcome technological barriers. These applications would have the cumulative effect of transforming the process into a more inclusive and deliberative space.

The main limitations to implementing new AI-based functionalities include: (1) challenges in integrating with the municipal administration's legacy systems; (2) cultural resistance to technological innovation in the public sector; and (3) risks of amplified digital exclusion if solutions are not accompanied by inclusion policies. The São Paulo case suggests that the most promising path lies in hybrid approaches, where AI acts as a facilitator—not a replacement—for human participation, creating conditions to enhance the deliberative nature that is essential to participatory budgeting and, in São Paulo's case, has only been partially employed.

However, the successful implementation of these solutions depends on developing specific state capacities. Three fundamental pillars stand out: (1) open data infrastructure and interoperable systems to feed algorithms; (2) multidisciplinary teams capable of mediating between technology and citizenship; and (3) robust algorithmic governance frameworks to ensure transparency and accountability. International experience suggests that the absence of these basic capacities may lead to the reproduction of biases or the technocratization of the participatory process.

These capacities shape both the feasibility and sustainability of implementing AI in participatory budgeting. While São Paulo's case suggests a solid institutional and technological base, analytical capacity remains a potential area for growth, particularly through AI tools that could assist in the preparation of the structure of proposals, highlight the main topics of the proposals by region or year, personalize engagement, or bring public data sources to assist in the development of proposals.